\documentclass[twocolumn,prl,showpacs,preprintnumbers,amsmath,amssymb]{revtex4}

\usepackage{graphicx}
\usepackage{dcolumn}
\usepackage{bm}

\begin{document}

\title{Pair Distribution Function of One-dimensional ``Hard Sphere'' Fermi and Bose Systems}

\author{Bo-Bo Wei$^1$}
\author{Chen-Ning Yang$^{1,2}$}

\affiliation{$^1${Chinese University of Hong Kong, Hong Kong}
$^2${Tsinghua University, Beijing, China}}

\begin{abstract}

The pair distributions of one-dimensional ``hard sphere'' fermion
and boson systems are exactly evaluated by introducing gap
variables.
\end{abstract}

\date{\today}
\maketitle



\section{Introduction}
Recent experimental progress in one-dimensional (1D) systems renews
theoretical interest in such systems. In this paper we study the
pair distribution function in 1D ``hard sphere'' systems.

Pair distribution function in 1D Tonks gas \cite{Tonks}and
Lieb-Liniger Bose gas \cite{Lieb1,Lieb2}  have been studied
previously, including the local correlation, $D(0)$, asymptotic
properties at large $r$, and the zero temperature
behavior\cite{ufford,Girardeau60,Girardeau65,Castin,Gangardt,Kheruntsyan1,Drummond,Astrakhsrchick,Cherny,Kheruntsyan2}.
The pair distribution function for 3D hard sphere boson gas has been
calculated by the authors \cite{bb}. In this work we address the
problem of pair distribution function in 1D hard sphere fermion and
boson systems with the diameter of the particles $a\neq 0$.

%

The pair distribution function is defined as
\begin{eqnarray}
D(r_{12})&=&\rho^{-2}\langle\psi^{\dag}(\textbf{r}_1)\psi^{\dag}(\textbf{r}_2)\psi(\textbf{r}_2)\psi(\textbf{r}_1)\rangle
\end{eqnarray}
where $\rho=N/L$ is the density. $D(r)$ is an important physical
quantity \emph{measurable} for many liquid systems.  In this formula
$\psi(\textbf{r})$ is the annihilation operator in $\textbf{r}$
space.

The pair distribution function $D(r_{12})$ is also related to the
\emph{diagonal elements} of the reduced density matrix
\cite{cnyang}:
\begin{eqnarray}
D(r_{12})=\rho^{-2}\text{Tr}[\psi(\textbf{r}_2)\psi(\textbf{r}_1)\rho_N\psi^{\dag}(\textbf{r}_1)\psi^{\dag}(\textbf{r}_2)],
\end{eqnarray}
where $\rho_N=\Psi_0 \Psi_0^{\dag}$,  and $\Psi_0$ is the many body
wave function.

The meaning of $D(r)$ is: given a particle A at one point, the
probability of finding another particle B at distance $r$
(counterclockwise) is
\begin{equation}
\rho D(r)dr.
\end{equation}
It is obvious that  $D(r)\rightarrow 1$ as $r\rightarrow \infty$.
also
\begin{equation}
\int_0^L \rho D(r)dr=N-1.
\end{equation}

\section{Fermions with $a=0$}
\subsection{Wave Function $\Psi$ For Fermions With $a=0$} First, we
consider $N$ free fermions in a one-dimensional cyclic interval of
length $L$. The momenta of these fermions are $2\pi k/L$, where
$k=-(N-1)/2$ to $(N-1)/2$. Here we consider the case that the number
of fermions in the system is odd, $N=2n+1$. The normalized many body
wave function $\Psi(r_1,r_2,\cdots,r_N)$ of the system is of the
form
\begin{eqnarray} \Psi=\frac{1}{\sqrt{N!}}\det
   \begin{pmatrix}
  \epsilon_1^n   & \epsilon_2^n  &\ldots& & \epsilon_{2n+1}^n \\
 \epsilon_1^{n-1}  & \epsilon_2^{n-1} &\ldots& & \epsilon_{2n+1}^{n-1}  \\
 \epsilon_1^{n-2}  & \epsilon_2^{n-2} &\ldots& & \epsilon_{2n+1}^{n-2}  \\
\vdots &\vdots  &\ddots& &\vdots  \\
\epsilon_1^{-n}  & \epsilon_2^{-n} &\ldots& & \epsilon_{2n+1}^{-n}
   \end{pmatrix}
\end{eqnarray}
where $\epsilon_j=\text{exp}({i \frac{2\pi}{L}r_j})/\sqrt{L}$. After
some calculations, we arrive at a compact form for the wave
function,
\begin{equation}
\Psi=\frac{1}{\sqrt{N!}}(\frac{1}{\sqrt{L}})^N
[2^{N(N-1)/2}]\prod_{1\leq i<j\leq N}\sin[\frac{\pi}{L}(r_j-r_i)]
\end{equation}
The pair distribution function $D(r)$ for free fermions can be
exactly obtained from this many-body wave function \cite{ufford}
\begin{equation}
D(r)=1-\frac{\sin^2(\frac{N\pi r}{L})}{N^2 \sin^2(\frac{\pi r}{L})}.
\end{equation}

\subsection{Gap Variables For Free Fermions} In order to prepare for
dealing with the $a\neq 0$ problem, we introduce gap variables in
the case $a=0$ as follows. Consider a region of the $N$ body
$L\times L\times \cdots \times L$ coordinate system where, modulo
$L$,
\begin{equation}
r_1 \leq r_2 \leq \cdots \leq r_N\leq r_1.
\end{equation}
We shall designate this region as $R_1$. In this region, we
introduce \emph{gap variables} $\{g_i\}$ :
\begin{equation}
g_i=r_{i+1}-r_i,\ \  i=1,2,3,\ldots,2n+1.
\end{equation}
Obviously $\sum g_i=L$. With the gap variables the many body wave
function $\Psi$ is given by
\begin{eqnarray}
\Psi&=&Q[f(g_1)f(g_2)\cdots f(g_{2n+1})] \nonumber\\
&&\times[f(g_1+g_2)f(g_2+g_3)\cdots f(g_{2n+1}+g_1)] \nonumber\\
&&\times[f(g_1+g_2+g_3)\cdots f(g_{2n+1}+g_1+g_2)] \nonumber\\
&&\times \cdots\cdots \nonumber \\
&&\times[f(g_1+\cdots+g_n)\cdots f(g_{2n+1}+g_1+\cdots+g_{n-1})]\nonumber\\
\end{eqnarray}
where $f(g)=\sin(\pi g/L)$, and $Q$ is the normalization factor
given by
\begin{equation}
Q=\frac{1}{\sqrt{N!}}(\frac{1}{\sqrt{L}})^N [2^{N(N-1)/2}].
\end{equation}
$\Psi$ has $(2n+1)\times n$ factors in total.

The probability distributions in $R_1$ is
\begin{equation}
|\Psi|^2 dg_1 dg_2\cdots dg_N \delta(\sum_{i=1}^N g_i-L)L
\end{equation}
with all $g_i\geq 0$. Now $R_1$ is only one of the $(N-1)!$ regions
of the full coordinate space. In each of these regions we have the
same gap distribution as (12). Thus gap distribution probability
$dP$ is
\begin{equation}
dP=(N-1)!L|\Psi|^2 dg_1 dg_2\cdots dg_N \delta(\sum_{i=1}^N g_i-L).
\end{equation}

\begin{figure}[h]
\begin{center}
\includegraphics[scale=0.4]{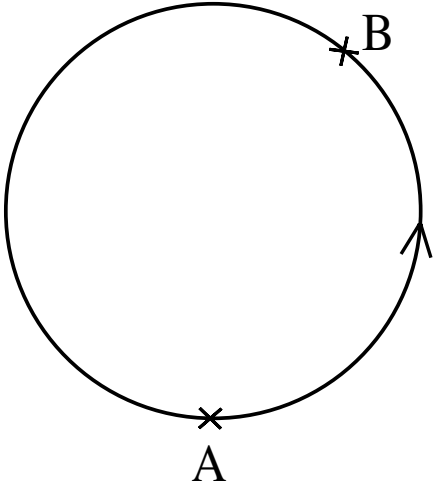}
\caption{\label{fig:epsart1}  Cycle of length $L$.}
\end{center}
\end{figure}

\subsection{Functions $F_i(r)$ For Free Fermions}
We now evaluate the function $D(r)$ of expression (3) using (13).
Going from particle A to B, counterclockwise in the cycle of length
$L$, as shown in Fig.~\ref{fig:epsart1},  there may be $i$ other
particles, with $i=0,1,\cdots,(N-2)$. Thus
\begin{equation}
\rho D(r)=\rho[F_0(r)+F_1(r)+\cdots+F_{N-2}(r)],
\end{equation}
where
\begin{equation}
\rho F_{i}(r)dr=\int \delta(\sum_{j=1}^{i+1}g_j-r)drdP
\end{equation}
Thus
\begin{eqnarray}
\rho F_{i}(r)&=&(N-1)!L\int|\Psi|^2 \delta(\sum_{j=1}^{i+1}g_j-r)
\nonumber\\
 &\times& \delta(\sum_{j=i+2}^{N}g_j-(L-r))dg_1dg_2\cdots dg_N
\end{eqnarray}
where all $g_j\geq 0$.

Integrating over $dr$ we get, from (15),
\begin{equation}
\int_0^L \rho F_i(r)dr=\int dP=1.
\end{equation}
Thus by (14),$\int \rho D(r)dr=N-1$, confirming (4). Outside of the
interval $(0,L)$ we define $F(r)$ by
\begin{equation}
F(r)=0 \ \ \text{for}\ \ r<0 \ \ \text{and} \ \ r>L.
\end{equation}

Besides (17), $F_i(r)$ has also the following properties:
\\(i) It is analytic except at $r=0$ and $r=L$, where $F(r)$ and its
first derivative are both zero.
\\(ii) $F_i(r)=F_{N-2-i}(L-r)$.

\begin{figure}[h]
\begin{center}
\includegraphics[scale=0.9]{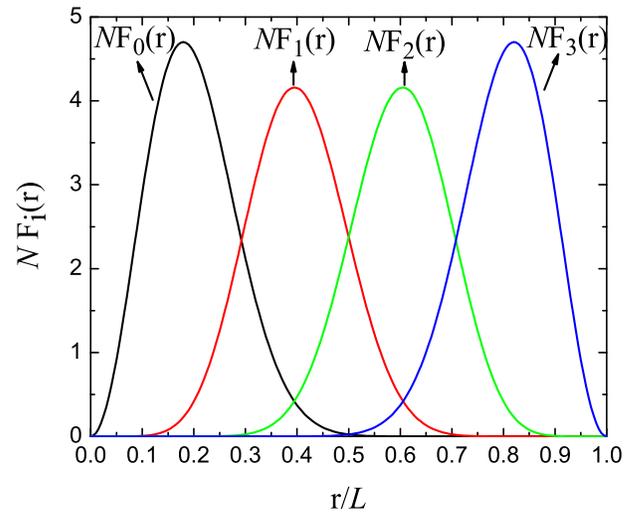}
\caption{\label{fig:epsart2} $N F_i(r)$ against $r/L$ for $N=5$ and
$i=0,1,2,3$.}
\end{center}
\end{figure}

It is obvious from (16) that $N F_i(r)$ is a function of $N$ and
$r/L$.  In Fig.~\ref{fig:epsart2} we plot this function vs $r/L$ for
the case $N=5$.

\begin{figure}[h]
\begin{center}
\includegraphics[scale=0.4]{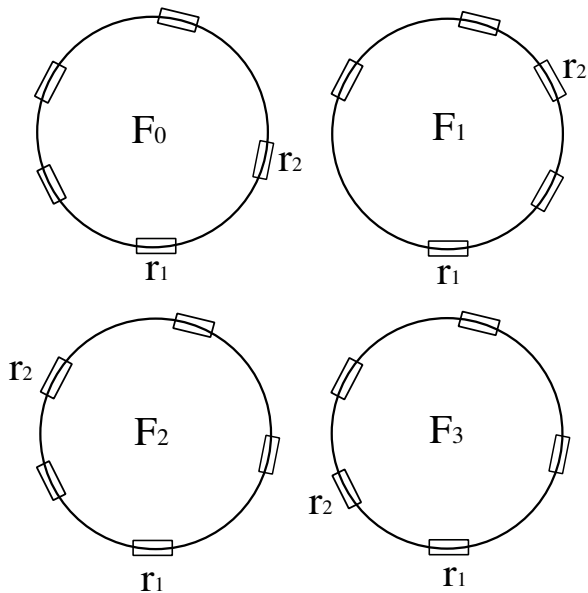}
\caption{\label{fig:epsart3}  4 possible configurations which
contribute to the pair distribution function $D(r)$ for $N=5$. The
ring represents the system and each block stands for a particle with
diameter $a$.}
\end{center}
\end{figure}

\begin{figure}[h]
\begin{center}
\includegraphics[scale=0.65]{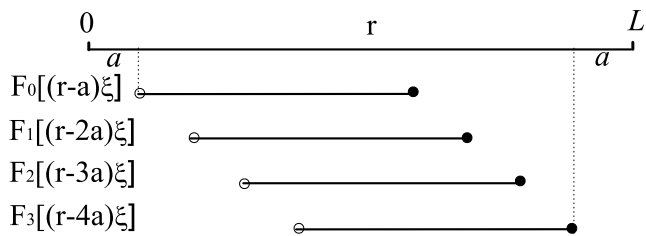}
\caption{\label{fig:epsart4} Regions where $F_i(r)>0$ for $N=5$.
Open circles are at $r=a,2a,3a,4a$. Closed circles are at
$r=L-a,L-2a,L-3a$ and $L-4a$. These $2(N-1)=8$ circles are where
$D(r)$ is singular. Each of the four regions has length $L-Na$.}
\end{center}
\end{figure}

\section{fermions with $a>0$}
For fermions with $a>0$, the full coordinate space is again divided
into $(N-1)!$ regions, one of which, $R_1$, is defined by the cyclic
condition (8) modulo $L$. We introduce gap variables in $R_1$ by
equations similar to (9):
\begin{eqnarray}
g_i&=&r_{i+1}-r_i-a, \ \ i=1,2,3,\cdots,(N-1) \\
\text{and}\ &&g_N=r_1-r_N+L-a  \hspace{3.3cm}(19a)\nonumber
\end{eqnarray}
Obviously
\begin{equation}
\sum g=L-Na.
\end{equation}

The key point is \emph{that $\Psi$ in terms of the gap variables
$\{g\}$ is still given} by (10), but with
\begin{equation}
f(g)=\sin(\pi g/(L-Na))
\end{equation}
and
\begin{equation}
Q=\frac{1}{\sqrt{N!}}(\frac{1}{\sqrt{L-Na}})^N [2^{N(N-1)/2}].
\end{equation}
Equation (12) now becomes
\begin{equation}
|\Psi|^2 dg_1 dg_2\cdots dg_N \delta(\sum_{i=1}^N g_i-(L-Na))L,
\end{equation}
and (13) beomes
\begin{equation}
dP=(N-1)!L |\Psi|^2 dg_1\cdots dg_N\delta(\sum_{i=1}^N g_i-(L-Na)).
\end{equation}

The function $D(r)$ is again, as in (14), a sum of $(N-1)F$
functions. To illustrate the reasoning we turn to
Fig.~\ref{fig:epsart3} and Fig.~\ref{fig:epsart4} for the case
$N=5$. We have
\begin{eqnarray}
D(r)&=& F_0[(r-a)\xi]+F_1[(r-2a)\xi] \nonumber \\
&+&F_2[(r-3a)\xi]+F_3[(r-4a)\xi],
\end{eqnarray}
where $\xi=L/(L-Na)$, and $F_i(r)$ is defined by (16) and (18).

Since $F_i(r)$ is nonzero only in the open interval $(0,L)$,
for $N=5$,\\$F_0[(r-a)\xi]$ is nonzero only for $a<r<L-4a$,\\
$F_1[(r-2a)\xi]$ is nonzero only for $2a<r<L-3a$,\\$F_2[(r-3a)\xi]$
is nonzero only for $3a<r<L-2a$,\\$F_3[(r-4a)\xi]$ is nonzero only
for $4a<r<L-a$. \\In Fig.~\ref{fig:epsart4} we indicate the regions
where $F_i[(r-(i+1)a)\xi]$ is nonzero for $i=0$ to 3.

\begin{figure}[h]
\begin{center}
\includegraphics[scale=0.9]{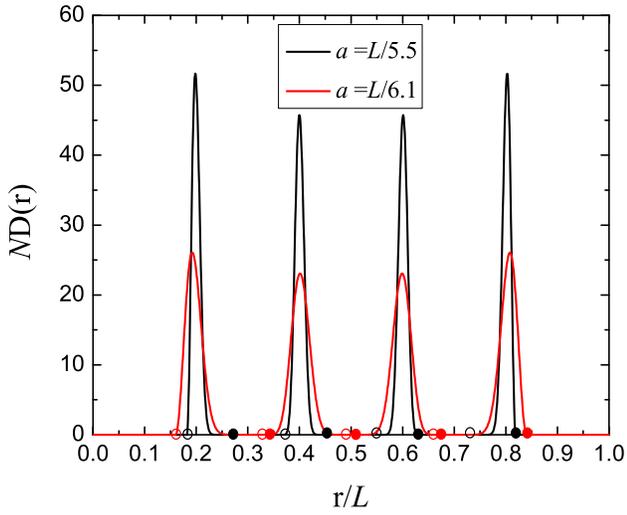}
\caption{\label{fig:epsart5}(color online)  The pair distribution
function $D(r)$ as a function of distance $r$ for $N=5$.  The
$N-1=4$ open circles and the $N-1=4$ closed circles are where $D(r)$
is singular. \\ \text{\hspace{0.4cm} }For case $a=L/(5+)$, $ND(r)$
is four $\delta$ functions (not shown).}
\end{center}
\end{figure}

\begin{figure}[h]
\begin{center}
\includegraphics[scale=0.9]{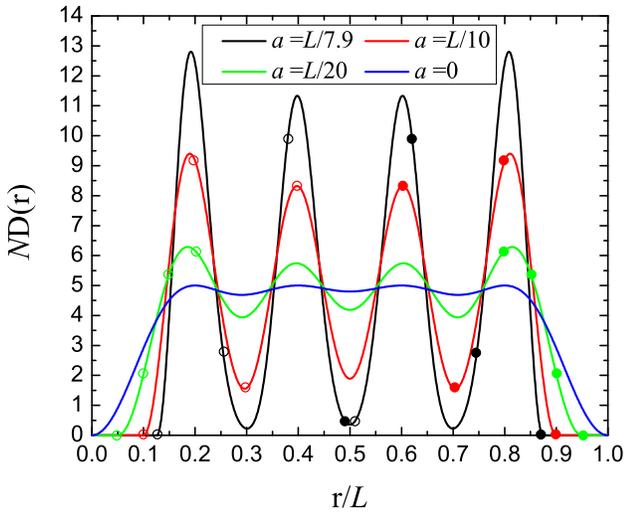}
\caption{\label{fig:epsart6}(color online)  The pair distribution
function $D(r)$ as a function of distance $r$ for $N=5$.}
\end{center}
\end{figure}

We present in Fig.~\ref{fig:epsart5} and Fig.~\ref{fig:epsart6} the
function $D(r)$ for $N=5$, and for several values of $L/a$.

Returning now to the general case, we conclude\\

\textbf{In the interval $(0,L)$, $D(r)$ is analytic everywhere
except at $(N-1)$ open circles at $r=a,2a,\cdots,(N-1)a$, and at
$(N-1)$ full circles at $r=L-a,L-2a,\cdots,L-(N-1)a$. At these
singular points $D(r)$ is continuous and has a continuous first
derivative.}

\section{Bosons with $a>0$}
For the one-dimensional hard sphere boson system, Girardeau
\cite{Girardeau60,Girardeau65} has shown a Bose-Fermi mapping
theorem which maps the hardcore boson system to a spinless hardcore
fermion system, for $a>0$.
\begin{equation}
\Psi_{\text{BE}}(x_1,x_2,\cdots,x_N)=|\Psi_{\text{FD}}(x_1,x_2,\cdots,x_N)|.\nonumber
\end{equation}
The pair distribution function is related to the square of the wave
function. Thus for $a>0$
\begin{equation}
D_{\text{BE}}(r)=D_{\text{FD}}(r).
\end{equation}

\end{document}